\documentclass[12pt,epsbox]{article}
\usepackage{latexsym}
\usepackage{psfig,amssymb,epsf}

\makeatletter
\@addtoreset{equation}{section}
\renewcommand{\theequation}{\thesection.\arabic{equation}}

\newcommand{\bC}{{\bf C}}

\newcommand{\bZ}{{\bf Z}}
\newcommand{\bN}{{\bf N}}

\newcommand{\bPhi}{\mbox{\boldmath $\Phi$}}
\newcommand{\bPsi}{\mbox{\boldmath $\Psi$}}
\newcommand{\bmZ}{\mbox{\boldmath $Z$}}
\newcommand{\bW}{\mbox{\boldmath $W$}}
\newcommand{\bV}{\mbox{\boldmath $V$}}

\newcommand{\cO}{{\cal O}}
\newcommand{\cU}{{\cal U}}

\newcommand{\nn}{\nonumber \\}

\newcommand{\be}{\begin{equation}} \newcommand{\ee}{\end{equation}}
\newcommand{\bea}{\begin{eqnarray}} \newcommand{\eea}{\end{eqnarray}}

\newcommand{\la}{\langle}
\newcommand{\ra}{\rangle}

\font\zfont = cmss10 
\newcommand{\ZZ}{\hbox{\zfont Z\kern-.4emZ}}

\ifx\TwoupWrites\UnDeFiNeD\else\target{\magstepminus1}{11.3in}{8.27in}
\source{\magstep0}{7.5in}{11.69in}\fi
\newfont{\fourteencp}{cmcsc10 scaled\magstep2}
\newfont{\titlefont}{cmbx10 scaled\magstep3}
\newfont{\authorfont}{cmcsc10 scaled\magstep1}
\newfont{\fourteenmib}{cmmib10 scaled\magstep2}
\skewchar\fourteenmib='177
\newfont{\elevenmib}{cmmib10 scaled\magstephalf}
\skewchar\elevenmib='177
\makeatletter
\newcommand\nonsequentialeqnum{
\@addtoreset{equation}{section}
\def\theequation{\arabic{section}.\arabic{equation}}}
\newif\ifp@bblock \p@bblocktrue
\newcommand\nopubblock{\p@bblockfalse}
\newcommand\topspace{\hrule height 0pt depth 0pt \vskip}
\newcommand\p@bblock{\begingroup \tabskip=\hsize minus \hsize
\baselineskip=1.5\ht\strutbox \topspace-2\baselineskip
\halign to\hsize{\strut ##\hfil\tabskip=0pt\crcr
\the\Pubnum\crcr\the\date\crcr}\endgroup}

\renewcommand\titlepage{\ifx\TwoupWrites\UnDeFiNeD\null
\vspace{-1.7cm}\fi
\vskip0.6cm
\ifp@bblock\p@bblock \else\hrule height 0pt \relax \fi}
\makeatother
\newtoks\date
\newtoks\Pubnum
\newtoks\pubnum
\newcommand{\frontpageskip}{\vspace{12pt plus .5fil minus 2pt}}
\renewcommand{\title}[1]{\frontpageskip
\begin{center}{\titlefont #1}\end{center}\par}
\renewcommand{\author}[1]{\frontpageskip\par\begin{center}
{\authorfont #1}\end{center}
\nobreak
}

\renewcommand{\thanks}[1]{\footnote{#1}}
\renewcommand{\abstract}{\par\frontpageskip\centerline{
\fourteencp Abstract}
\vspace{8pt plus 3pt minus 3pt}}

\begin{document}

\begin{titlepage}
\hfill
\vbox{
    \halign{#\hfil         \cr
           CERN-TH/2002-188 \cr
           TAUP-2711-02 \cr
           hep-th/0208078  \cr
           } 
      }  
\vspace*{20mm}
\begin{center}
{\Large {\bf   Exact Anomalous Dimensions for
${\cal N}=2$ ADE SCFTs}\\} 
\vspace*{15mm}
{\sc Yaron Oz}$^{a\,b}$ 
\footnote{e-mail: {\tt yaronoz@post.tau.ac.il, Yaron.Oz@cern.ch}}
and {\sc Tadakatsu Sakai}$^{a}$
\footnote{e-mail: {\tt tsakai@post.tau.ac.il}}

\vspace*{1cm} 
{\it {$^{a}$ Raymond and Beverly Sackler Faculty of Exact Sciences\\
School of Physics and Astronomy\\
Tel-Aviv University , Ramat-Aviv 69978, Israel}}\\ 

\vspace*{5mm}
{\it {$^{b}$Theory Division, CERN \\
CH-1211 Geneva  23, Switzerland}}\\

\end{center}

\begin{abstract}

We consider four-dimensional
${\cal N}=2$ superconformal field theories based on $ADE$ quiver diagrams.
We use the  procedure of hep-th/0206079 and
compute the exact anomalous dimensions of operators 
with large $U(1)_R$ charge to all orders in perturbation
in the planar limit.
The results are in
agreement with the string
computation in the dual pp-wave backgrounds.

\end{abstract}
\vskip 2cm

August 2002

\end{titlepage}

\setcounter{footnote}{0}

\newpage

\section{Introduction}

The pp-wave background 
has received much attention recently since 
it can
be obtained by taking a Penrose limit of AdS$_5\times S^5$ \cite{bmn}.
Via the AdS/CFT correspondence (for a review, see \cite{ads}),
it corresponds to a particular limit of  ${\cal N}=4$ SYM theory, where 
the number of colors $N$ is taken to infinity, the Yang-Mills coupling $g_{YM}$ is
kept fixed and one considers operators with infinite R-charge $J$ such
that $J/\sqrt{N}$ is fixed.
Since string theory on a pp-wave background is solvable,
it was argued in \cite{bmn} that
the anomalous dimensions of a class of operators with large R-charge
can be predicted to all orders in perturbation in the planar limit.

Obviously, it would be nice if the string theory predictions
can be confirmed on the CFT side.
For ${\cal N}=4$ SYM theory, the authors of \cite{zanon} performed
an exact computation of the anomalous dimensions.
The results were in agreement with
the string theory analysis.

The aim of this letter is to perform a similar computation
for ${\cal N}=2$ 
superconformal field theories (SCFTs) based on $ADE$ quiver diagrams.
These theories are conjectured to be dual  to IIB string on
AdS$_5\times S^5/\Gamma$ \cite{ks} where
$\Gamma$ is one of the $ADE$ finite subgroup of $SU(2)$.

The letter is organized as follows.
In section 2 we describe the construction of the ${\cal N}=2$ quiver
gauge theories.
In section 3 we compute the exact spectrum of anomalous dimensions.

\section{Quiver gauge theories}

The quiver gauge theories can be constructed by considering $N|\Gamma(G)|$
D3-branes on the covering space of $\bC^2/\Gamma(G)$ and performing
a $\Gamma(G)$ projection on the worldvolume fields and the Chan-Paton
factors \cite{dm,jm,lnv}.
We start with 
${\cal N}=4$ SYM theory with the gauge group $SU(N|\Gamma(G)|)$
and the action given by
\begin{eqnarray}
S&\!=\!& \int d^4x\,d^4\theta~ {\rm tr}\left(e^{-g_{\rm YM}\bV}
\bar{\bPhi}_i\, e^{g_{\rm YM}\bV} \bPhi^i\right) \nn
&&+\,\frac{1}{2g_{\rm YM}^2}
\int d^4x\,d^2\theta \,{\rm tr}\left(\bW^\alpha \bW_\alpha 
\right)
+\frac{ig_{\rm YM}}{3!} \int d^4x\,d^2\theta\, \epsilon_{ijk}
\,{\rm tr}\left( \bPhi^i [\bPhi^j,\bPhi^k]\right)+{\rm c.c.}
\label{n=4action}
\end{eqnarray}
The ${\cal N}=1$ vector superfield $\bV$ and chiral superfields
$\bPhi^i,~i=1,2,3$ belong to the adjoint
representation of $SU(N|\Gamma(G)|)$.
$g_{\rm YM}^2=4\pi g_s$ is the gauge coupling of the ${\cal N}=4$
SYM.
We follow the notations adopted in \cite{zanon}.
The manifest global symmetries of this action are 
\begin{eqnarray}
U(1)_X:&&~~ 
\bW(\theta) \rightarrow e^{i\alpha}\bW(e^{-i\alpha}\theta),~
\bPhi^i(\theta)\rightarrow e^{i{2\over 3}\alpha}\bPhi^i(e^{-i\alpha}\theta),
\nn
&&\cr
SU(3):&&~~
\bW(\theta) \rightarrow \bW(\theta),~
\bPhi^i(\theta)\rightarrow U^i_j\,\bPhi^j(\theta) \ .
\end{eqnarray}
We denote
$\bPhi\equiv\bPhi^1,~\bPsi\equiv\bPhi^2,\bmZ\equiv\bPhi^3$.

The orbifold group $\Gamma(G)$ is a subgroup
\begin{equation}
\Gamma(G)\subset SU(2)\subset SU(3) \ .
\label{group}
\end{equation}
By this orbifolding, ${\cal N}=4$ SUSY is broken to
${\cal N}=2$ SUSY where the $U(1)_R$ symmetry is an
appropriate linear combination of $U(1)_X$ and $U(1)^{\prime}$.
$U(1)^{\prime}$ is the normal subgroup to 
$SU(2)\subset SU(3)$.
There is also an $SU(2)_R$ symmetry under which 
the scalar components of $\bPhi,\bar{\bPsi}$ transform as a doublet.
The orbifolding for the worldvolume fields is implemented by
the conditions
\begin{equation}
R\bW R^{-1}=\bW,~~R\bmZ R^{-1}=\bmZ,~~
\left(
\begin{array}{c}
R\bPhi R^{-1} \nn
R\bPsi R^{-1}
\end{array}
\right)
=
Q
\left(
\begin{array}{c}
\bPhi \nn
\bPsi
\end{array}
\right).
\label{projection}
\end{equation}
$R$ is the $N|\Gamma(G)|$-dimensional regular
representation of $\Gamma(G)$ which can be decomposed as
$\oplus_{i=1}^r(n_ir_i\otimes I_N)$ with $r_i$ being the irreducible
representations of $\Gamma(G)$. $r={\rm rank}G$, and 
$Q$ is a two-dimensional representation of $\Gamma(G)$.

\subsection{$A_{k-1}$ orbifolds}

Consider $\Gamma(A_{k-1})=\bZ_k$.
$R$ can be taken to be 
${\rm diag}(1,\omega^{-1},\omega^{-2},\cdots,\omega^{-k+1})$
where $\omega=e^{2\pi i/k}$ and each block is proportional to
an $N\times N$ unit matrix.
$Q$ is the $2\times 2$ matrix of the form 
${\rm diag}(\omega,\omega^{-1})$.
It is easy to verify that the following components survive the
projections
\begin{eqnarray}
&&\bW=\left(
\begin{array}{cccc}
W_0 &     &        & \\
    & W_1 &        & \\
    &     & \ddots & \\
    &     &        & W_{k-1}
\end{array}
\right)~~~~
\bmZ=\left(
\begin{array}{cccc}
Z_0 &     &        & \\
    & Z_1 &        & \\
    &     & \ddots & \\
    &     &        & Z_{k-1}
\end{array}
\right) \nn
&&\cr
&&\bPhi=\left(
\begin{array}{cccc}
0         & \Phi_{01} &                 &        \\
          & 0         & \Phi_{12}       &        \\
          &           & \ddots          &  \ddots \\
\Phi_{k-1,0} &           &                 &  
\end{array}
\right) ~~~~
\bPsi=\left(
\begin{array}{cccc}
0         &           &                 &  \Psi_{k-1,0}      \\
\Psi_{01} & 0         &                 &                 \\
          & \Psi_{12} & \ddots          &                 \\
          &           & \ddots          &  0
\end{array}
\right)
\end{eqnarray}
The resulting theory is an $\prod_{i=0}^{k-1} SU(N)_i$ gauge theory with
the ${\cal N}=1$ chiral superfields  $\Phi_{i,i+1}$ in
the $({\bN,\bar{\bN}})$ 
representation of
$SU(N)_i\times SU(N)_{i+1}$, and $\Psi_{i,i+1}$ in
the $({\bar{\bN},\bN})$ representation of $SU(N)_i\times SU(N)_{i+1}$ 
\footnote{The sub-index $k$ is equivalent to $0$.}.

\subsection{$D,E$ orbifolds}

$\Gamma(D_k)={\bf D}_{k-2}$ is the binary extension of the dihedral
group of order $4(k-2)$.
$\Gamma(E_6)={\cal T}$, $\Gamma(E_7)={\cal O}$ and
$\Gamma(E_8)={\cal I}$ are the binary tetrahedral, octahedral and
icosahedral groups of order 24, 48 and 120, respectively.
The gauge group and matter content of the $D$ and $E$ theories are
worked out by
the same procedure. 
The results are summarized
in terms of quiver diagrams corresponding to the extended Dynkin diagrams
of the group.
The gauge group is associated with the nodes
of the diagram. It is
$\prod_{i=0}^r G_i$ with $G_i=SU(n_iN)$  where 
$n_i$ is the dimension of the irreducible representation
of $\Gamma(G)$ corresponding to the node $i$.
The matter content is associated with the links of the diagram.
A link between the node $i$ and $j$ is
a hypermultiplet
$H_{\la ij \ra}=(\Phi_{\la ij\ra},\bar{\Psi}_{\la ij\ra})$
which transforms in the bifundamental representation
of $G_i\times G_j$. 
The ${\cal N}=1$ chiral superfields in the adjoint representation
of $\prod_i G_i$ which are part of
the ${\cal N}=2$ vector multiplets
are denoted by $Z_i$.
Recall that the $U(1)_R$ charge of $\bmZ$ is 2 while $\bPhi$ and
$\bPsi$ are neutral.

\section{Anomalous dimensions}

The action of the quiver gauge theories is given by (\ref{n=4action}) 
divided by $|\Gamma(G)|$ with only 
the fields that remain after the projections (\ref{projection}).
The field equations read
\begin{eqnarray}
&&\bar{D}^2\bar{\bPhi}+ig_{\rm YM}(\bPsi\bmZ-\bmZ\bPsi)=0,\nn
&&\bar{D}^2\bar{\bPsi}+ig_{\rm YM}(\bmZ\bPsi-\bPsi\bmZ)=0,\nn
&&\bar{D}^2\bar{\bmZ}+ig_{\rm YM}(\bPhi\bPsi-\bPsi\bPhi)=0 \ .
\label{eoms}
\end{eqnarray}

The propagators take the form
\begin{eqnarray}
&&\langle Z_i(z)\bar{Z}_j(z^{\prime})\rangle
=\delta_{ij}\,{|\Gamma(G)| \over 4\pi^2n_i}\,\bar{D}^2
{\delta^4 (\theta-\theta^{\prime})\over |x-x^{\prime}|^2}
\overleftarrow{D}^2, \nn
&&\langle \Phi_{\la ij\ra}(z)\bar{\Phi}_{\la i^{\prime}j^{\prime}\ra}
(z^{\prime})\rangle
=\langle \Psi_{\la ij \ra}(z)\bar{\Psi}_{\la i^{\prime}j^{\prime}\ra}
(z^{\prime})\rangle 
=c_{\la ij\ra}\delta_{ij}\delta_{i^{\prime}j^{\prime}}\,
{|\Gamma(G)| \over 4\pi^2}\,\bar{D}^2
{\delta^4 (\theta-\theta^{\prime})\over |x-x^{\prime}|^2}
\overleftarrow{D}^2.
\label{propagator}
\end{eqnarray}
$c_{\la ij\ra}$ are constants that are fixed by the action.
For $G=A_{k-1}$ 
$c_{\la ij\ra}=1$ for all $\la ij\ra$.
For $G=D_4$ $c_{\la ij\ra}=1/2$.
We will not work out $c_{\la ij\ra}$ in general
as it is not needed in the sequel.

Consider now the dictionary between CFT operators and string states
on the pp-wave background \cite{ali,kim,taka}.
\footnote{For an earlier work on comparison of the spectrum between
AdS orbifolds and the dual ${\cal N}=2$ quiver gauge theories, 
see \cite{oz}\cite{gukov}.}
We will work in the light-cone gauge. 
Let us discuss first the ground states $|0,2p^+\rangle_q$, where
$q=0,1,\cdots, r$ label the $q$-twisted sectors of the
string theory.
These ground states can be identified with the CFT operators
${\rm tr}(R_q\bmZ^J)$. 
$R_q$ are representatives of the conjugacy classes of the 
$N|\Gamma(G)|$ dimensional regular
representation of $\Gamma(G)$.
$R_0$ corresponds to the identity.
Recall that the operators depend only on the conjugacy classes
by using an appropriate constant gauge transformation
of $SU(N|\Gamma(G)|)$.
Now consider the CFT operators
\begin{eqnarray}
&&{\cal O}_J=
\left({|\Gamma(G)|\over 4\pi^2}\right)^{-(J+1)/2}\!N^{-J/2}
\sum_l e^{i\varphi(q) l}R_q\bmZ^l\bPhi\bmZ^{J-l}, \nn
&&\cr
&&{\cal U}_J=
\left({|\Gamma(G)|\over 4\pi^2}\right)^{-(J+1)/2}\!N^{-J/2}
\sum_l e^{i\varphi(q) l}R_q\bmZ^l\bar{\bPsi}\bmZ^{J-l},
\end{eqnarray}
where $\varphi(q)=2\pi n(q)/J$. 
The overall normalization factors are chosen in order to get later a finite result
for the two-point functions in the large $N$ and $J$ limit.
The string state corresponding to ${\cal O}_J$ is 
obtained by acting with a creation operator in the $\bC^2/\Gamma(G)$ part
with the oscillation number $n(q)$. 
Note that we are working in the ``dilute gas'' approximation
\cite{bmn,gross} that is valid for $J\gg 1$.
This implies that the diagrams relevant to the anomalous dimension
of an operator with a large number of impurities involve only one
impurity and $\bmZ$ fields next to the impurity field.
The operators ${\cal O}_J,{\cal U}_J$ are the building
blocks for it.

We are interested in the exact form of the anomalous dimension
of ${\cal O}_J$, $\gamma$. 
To compute this, we first notice that (\ref{eoms}) implies
the relation 
\begin{equation}
\la\bar{D}^2{\cal{U}}_J(z)D^2\bar{{\cal{U}}}_J(z^{\prime})\ra
=-{g_{\rm YM}^2N|\Gamma (G)|\over 4\pi^2}\,\alpha(q)\, \la{\cal{O}}_{J+1}(z)
\bar{{\cal{O}}}_{J+1}(z^{\prime})\ra \ ,
\label{22pt}
\end{equation}
with $z=(x,\theta,\bar{\theta})$ being the ${\cal N}=1$ superspace
coordinates and
$\alpha(q)=-(e^{i\varphi(q)}-1)(e^{-i\varphi(q)}-1)$.
The computation of $\gamma$ proceeds exactly the same way as in the
case of ${\cal N}=4$ SYM \cite{zanon}.
Let us first compute the two-point functions of ${\cal O}_{J+1}$ and
${\cal U}_J$.
Notice that ${\cal O}_{J+1}$ and ${\cal U}_J$ consist
of the blocks of the form 
\begin{eqnarray}
({\cal O}_{J+1})_{\la ij \ra}&\!\equiv\!&
\left({|\Gamma(G)|\over 4\pi^2}\right)^{-(J+2)/2}\!N^{-(J+1)/2}
\sum_l e^{i\varphi(q)l}(Z_i)^l\,\Phi_{\la ij\ra}\,(Z_j)^{J-l+1}, \nn
({\cal U}_{J})_{\la ij \ra}&\!\equiv\!&
\left({|\Gamma(G)|\over 4\pi^2}\right)^{-(J+1)/2}\!N^{-J/2}
\sum_l e^{i\varphi(q)l}(Z_i)^l\,\bar{\Psi}_{\la ij\ra}\,(Z_j)^{J-l}.
\end{eqnarray}
Using the propagators (\ref{propagator}), the two-point functions at
tree level read 
\begin{eqnarray}
\la({\cal{O}}_{J+1})_{\la ij \ra}(z)
\,(\bar{{\cal{O}}}_{J+1})_{\la ij \ra}(z^{\prime})\ra &\!\!=\!\!&
c_{\la ij \ra} 
\left(\bar{D}^2D^2
-{i\over 4}{\Delta_{\cO}-\omega_{\cO} \over \Delta_{\cO}}
\left[ D^{\alpha},\bar{D}^{\dot{\alpha}}\right]\partial_{\alpha\dot{\alpha}}
\right. \nn 
&&~~~~~~~~~~~~~~~\left.
-{1\over 4}{(\Delta_{\cO}+\omega_{\cO})(\Delta_{\cO}-\omega_{\cO})
\over \Delta_{\cO}(\Delta_{\cO}-1)}\,\Box\right)
\frac{\delta^4(\theta-\theta^{\prime})}
{|x-x^{\prime}|^{2\Delta_{\cO}}}, \nn
\la\bar{D}^2({\cal{U}}_J)_{\la ij \ra}(z)\,
D^2(\bar{{\cal{U}}}_J)_{\la ij \ra}(z^{\prime})\ra &\!\!=\!\!&
(\Delta_{\cU}-\omega_{\cU})(\Delta_{\cU}-\omega_{\cU}-2)
\,c_{\la ij \ra}\,\bar{D}^2 D^2
\frac{\delta^4(\theta-\theta^{\prime})}{|x-x^{\prime}|^{2(\Delta_{\cU}+1)}} \ .
\end{eqnarray}
Here $\Delta_{\cO,\cU}$ and $\omega_{\cO,\cU}$ are the scaling dimensions
and the chiral weights of $\cO_{J+1},\cU_J$ and given by
$\Delta=h+\bar{h},\omega=h-\bar{h}$ with $h,\bar{h}$ being the number of
chiral and anti-chiral superfields, respectively.

The two-point functions receive quantum corrections: the scaling
dimensions shift by the anomalous dimensions 
$\Delta_{\cU,\cO}\rightarrow\Delta_{\cU,\cO}+\gamma_{\cU,\cO}$,
and the overall factor $c_{\la ij \ra}$ should be replaced by
some functions of the coupling constant.
Furthermore the chiral weights may change as 
$\omega_{\cU,\cO}\rightarrow\omega_{\cU,\cO}+\delta\omega_{\cU,\cO}$.
It then follows that the exact two-point functions in the planar and
large $J$ limit take the form
\begin{eqnarray}
\la\bar{D}^2({\cal{U}}_J)_{\la ij \ra}(z)\,
D^2(\bar{{\cal{U}}}_J)_{\la ij \ra}(z^{\prime})\ra &\!\!=\!\!&
(\gamma_{\cU}-\delta\omega_{\cU})(\gamma_{\cU}-\delta\omega_{\cU}-2)\,
f_{\la ij \ra}^{\cU} 
\bar{D}^2 D^2
\frac{\delta^4(\theta-\theta^{\prime})}{|x-x^{\prime}|^{2(J+2+\gamma_{\cU})}}, \nn
&&\cr
\la({\cal{O}}_{J+1})_{\la ij \ra}(z)
\,(\bar{{\cal{O}}}_{J+1})_{\la ij \ra}(z^{\prime})\ra &\!\!=\!\!&
f_{\la ij \ra}^{\cO}\,\bar{D}^2D^2
\frac{\delta^4(\theta-\theta^{\prime})}{|x-x^{\prime}|^{2(J+2+\gamma_{\cO})}}
\ ,
\end{eqnarray}
with $f_{\la ij \ra}^{\cU,\cO}=c_{\la ij \ra}+O(g_{\rm YM}^2N/J^2)$.
Here we assume that $\gamma$ and $\delta\omega$ are of order one.
We now observe that 
\begin{equation}
\gamma_{\cU}=\gamma_{\cO}\equiv\gamma,~~
\delta\omega_{\cU}=\delta\omega_{\cO}\equiv\delta\omega,~~
f_{\la ij \ra}^{\cU}=f_{\la ij \ra}^{\cO}\equiv f_{\la ij \ra} \ .
\end{equation}
To see this, 
recall that ${\cal{O}}_{J+1}$ differs from ${\cal{O}}_{J}$ only by
the presence of an extra $\bmZ$-field. 
As pointed out in \cite{zanon},
this difference is irrelevant to the renormalization
in the dilute gas approximation.
Also since the scalar components of ${\cal O}_J$ and ${\cal U}_J$
belong to the same ${\cal N}=2$ multiplet, one can use $SU(2)_R$
symmetry to see that
$\la{\cal{U}}_{J}(z)\,\bar{{\cal{U}}}_{J}(z^{\prime})\ra=
\la{\cal{O}}_{J}(z)\,\bar{{\cal{O}}}_{J}(z^{\prime})\ra$.

Plugging into (\ref{22pt}), we obtain
\begin{equation}
(\gamma-\delta\omega)(\gamma-\delta\omega-2)
=-{g_{\rm YM}^2N|\Gamma(G)|\over 4\pi^2}\,\alpha(q) \ .
\label{rel;gamma}
\end{equation}
Under the assumption that $\delta\omega=0$, the anomalous dimension
in the large $J$ limit can be solved to be
\begin{equation}
\gamma=-1+\sqrt{1+{g_{\rm YM}^2N|\Gamma(G)|\over J^2}\,n(q)^2} \ .
\label{gamma}
\end{equation}

When setting $g_{\rm YM}^2=4\pi g_s$, this result
coincides with the string
computation by choosing $n(q)$ to be the oscillation number.


\noindent
\underline{$G=A_{k-1}$}

For $G=A_{k-1}$, one has $n(q)=n+{q\over k}$ with
$n\in\bZ$ and $q=0,1,\cdots, k-1$.

\noindent
\underline{$G=D_{k}$}

The $4(k-2)$ elements of the two-dimensional defining representation 
of $\Gamma(D_k)$ consist of the matrices of the form
\footnote{See, for instance, \cite{ale}.}
\begin{equation}
F_l=\left(
\begin{array}{cc}
e^{i\pi l\over k-2} & 0    \nn
0 & e^{-{i\pi l\over k-2}}
\end{array}
\right), \quad
G_l=\left(
\begin{array}{cc}
0 & ie^{-{i\pi l\over k-2}}    \nn
ie^{i\pi l\over k-2} & 0
\end{array}
\right) \ ,
\end{equation}
with $l=0,1,\cdots, 2k-5$.
There are $k+1$ conjugacy classes:
\begin{itemize}
\item $KE$ contains only the identity $F_0$.

\item $KZ$ contains the central extension ${\cal Z}\equiv F_{k-2}=-1$.

\item $KG_{\rm e}$ contains the elements $G_{2\nu}$ with
      $\nu=1,2,\cdots, k-3$.

\item $KG_{\rm o}$ contains the elements $G_{2\nu+1}$.

\item There exist $k-3$ classes denoted by 
      $KF_{\mu},~\mu=0,1,\cdots,k-3$. Each of these contains the pair of
      the elements $\{ F_{\mu},F_{2(k-2)-\mu} \}$.
\end{itemize}
The matrices in each conjugacy class set
a boundary condition for the 2d fields of the orbifold CFT on
$\bC^2/\Gamma(D_k)$.
One can verify that the resulting (un)twisted sectors contain
string modes with the following oscillation numbers
\begin{eqnarray}
n(KE)=n,~~n(KZ)=n+{1\over 2},~~n(KG_{\rm e})=n(KG_{\rm o})=n+{1\over 4},~~
n(KF_{\mu})=n+{\mu\over 2(k-2)}.
\end{eqnarray}

One can work out in a similar way the explicit oscillation numbers
for $G=E_{k}$.

\section*{Acknowledgements}

This research is supported by the US-Israel Binational Science
Foundation.



\end{document}